\documentclass{elsart}

\usepackage{amssymb}
\usepackage{graphicx}
\usepackage{units}
\usepackage{color}
\usepackage{url}

\begin{document}

\newcommand{\pt}{p_\perp}
\newcommand{\kt}{k_\perp}
\newcommand{\tn}[1]{\mathrm{#1}}
\newcommand{\df}{\Delta \phi}
\newcommand{\mean}[1]{\langle #1 \rangle}

\begin{frontmatter}

\title{Jet Quenching from soft QCD Scattering\\ in the Quark-Gluon Plasma}

\author[hd]{Korinna Zapp}, 
\author[gi]{Gunnar Ingelman},
\author[gi]{Johan Rathsman},
\author[hd]{Johanna Stachel}

\address[hd]{Physikalisches Institut, Universit\"at Heidelberg,
Philosophenweg 12, D-69120 Heidelberg, Germany} 

\address[gi]{High Energy Physics, Uppsala University, Box 535, S-75121
Uppsala, Sweden}

\begin{abstract}
We show that partons traversing a quark-gluon plasma can lose substantial
amounts of energy also by scatterings, and not only through medium-induced
radiation as mainly considered previously. Results from Monte Carlo simulations
of soft interactions of partons, emerging from a hard scattering, through
multiple elastic scatterings on gluons in an expanding relativistic plasma show
a sizeable jet quenching which can account for a substantial part of the effect
observed in RHIC data.
\end{abstract}

\end{frontmatter}

Partons emerging from hard scattering processes and traversing a quark-gluon
plasma (QGP) are expected to lose a substantial amount of their energy through
interactions with the plasma, resulting in a suppression of high-$\pt$
particles \cite{Wang:1991xy,Baier:1996kr}.  Such jet quenching has indeed been
observed at RHIC  \cite{Arsene:2004fa,Back:2004je,Adams:2005dq,Adcox:2004mh} 
and is believed to be not only a signal for QGP formation, but also a tool for
investigating the properties of the plasma. So far theoretical efforts have
concentrated on medium-induced gluon bremsstrahlung as energy loss mechanism
\cite{Wang:2003mm,Wang:2004tt,Gyulassy:2000er,Wiedemann:2000za,Salgado:2003gb,Zakharov:1996fv}
based on arguments that the collisional energy loss is small
\cite{Baier:2000mf,Thoma:1990fm,Thoma:1995ju}. In this letter, however, we
demonstrate that energy loss caused by multiple elastic scatterings is not
negligible, but can contribute significantly to the observed jet quenching
effect. The importance was also realised in \cite{Mustafa:2003vh,Dutt-Mazumder:2004xk}
and scattering was included at the same level as radiation in
\cite{Lokhtin:2005px}.

Such multiple scatterings can be related to recent developments to understand
soft QCD interactions of a high energy parton with a colour background field.
This has, in particular, been used as a novel way to understand diffractive
hard scattering in $ep$ or $p\bar{p}$ collisions \cite{Ingelman:2005ku}. 
In the Soft Colour Interaction (SCI) model \cite{Edin:1995gi,Enberg:2001vq} 
a hard-scattered parton interacts with the proton remnant via soft gluon
exchanges. Here, it is the exchange of colour charge which is important, since
this changes the colour topology of the event such that hadronisation will
produce a different final state, e.g.\ diffractive ones with a gap in the
rapidity distribution of hadrons. The phenomenological success of the SCI model
indicates that it captures the most essential QCD dynamics for soft gluon
exchanges. A theoretical basis for this model has recently been found
\cite{Brodsky:2004hi} in terms of QCD rescatterings of an energetic parton with
the target colour field via one or more gluons as expressed through the Wilson
line in the parton density definition. 

In this letter, we develop the SCI model to apply for a parton that
rescatters in a quark-gluon plasma. The essential quantity for the energy loss
through such elastic scattering is then the energy-momentum transfer involved, 
rather than the colour exchange. Since the QGP is much denser and of larger
volume than the colour background field of a single proton, the parton should
experience many more interactions so that even a small momentum transfer per
scattering could add up and result in a sizeable energy loss.

The SCI jet quenching model \cite{Zapp:2005wr} is implemented as a Monte Carlo
event generator based on \textsc{Pythia} (version 6.2)
\cite{Sjostrand:2000wi} which is used to simulate hard scattering processes
based on perturbative QCD $2\to 2$ matrix elements and initial and final state
parton showers based on DGLAP evolution. These perturbative QCD (pQCD)
processes are not altered, but are used as in standard high-$p_\perp$ $pp$
collisions. However, before the emerging partons hadronise, we introduce the
possibility that they may interact with the plasma. 

This interaction of the hard parton with the plasma is treated as elastic
scattering on the partons in the QGP, with a squared momentum transfer $t$.
Although this is treated as a $2\to 2$ parton-parton scattering (in the cm
frame), we cannot use pQCD matrix elements since that would only apply for the
small cross-section processes with large $t$. Instead we want to examine energy
loss through the dominant soft interactions where perturbation theory does not
apply. Being unable to calculate a proper $t$-distribution from first
principles, we assume that it can be approximated by a Gaussian distribution
having a width $\sigma_t$ as a parameter of the model. Naturally, the mean
momentum transfer should not exceed a few hundred \unit{MeV} in order to be
non-perturbative. 

The partons emerging from the hard pQCD phase are traced through the plasma and
scatter, with a probability $P_\tn{int}$, on each plasma gluon along their way
within a screening radius $R_\tn{scr}$. We use $P_\tn{int}\simeq 0.5$ since
this was found as a universal parameter value for the SCI model to fit rapidity
gap data from both $ep$ and $p\bar{p}$ collisions \cite{Edin:1995gi,Enberg:2001vq}, 
as well as charmonium formation \cite{SCI-charmonium,Mariotto:2001sv}. 

The QGP is represented by an ideal relativistic gluon gas, while quarks are
neglected because the QGP is initially gluon rich and the gluons come much
faster into thermal equilibrium than the quarks. This means that the number and
energy densities of gluons are connected to the temperature via $n_g = g_g\, T^3
2\,\zeta(3)/2\pi^2$ and $\epsilon_g = \pi^2 g_g\,T^4/30$ so that $ n_g \propto
\epsilon_g^{3/4}$.

The time evolution is governed by a Bjorken-like longitudinal expansion
\cite{Bjorken:1982qr}, i.e.\ with the equation of state of an
ultra-relativistic ideal gas the time dependence of the energy density and
temperature is given by $\epsilon(\tau) \propto \tau^{-4/3}$ and $T(\tau)
\propto \tau^{-1/3}$,  where $\tau = \sqrt{t^2-z^2}$ is the proper time.
Therefore, the density of gluons drops very fast, namely as $n_g(\tau) \propto
\tau^{-1}$. It is assumed that the local initial energy density is proportional
to the number of binary nucleon-nucleon collisions at impact parameter $b$,
i.e.\ $\epsilon(x,y,b)\propto T_\mathrm{Au}(x-b/2,y)\cdot
T_\mathrm{Au}(x+b/2,y)$ where the nuclear thickness function $T_\mathrm{Au}$ is
estimated with a simple Glauber model \cite{Eskola:1988yh}. 

The parameter that governs the energy density of the plasma is $\epsilon_0$,
the energy density in most central collisions (i.e.\ $b=0$) at $\tau_0 =
\unit[1]{fm/c}$ averaged over the transverse area. This fixes the normalisation
of the density profile for any centrality.

The centrality of a nucleus-nucleus collision is defined by the fraction
of the total geometrical cross section it takes. A centrality class, i.e.\ a
range in centrality can be translated into an impact parameter range using
the Glauber model calculation. In the simulation an impact parameter is chosen
for each event in a given range of centrality according to $\d \sigma
\propto b\d b$ (ignoring the fluctuations in the experimental
quantity used to determine the centrality).

The hard scatterings are treated as in $pp$ collisions and are
distributed in the transverse plane according to the number of binary
nucleon-nucleon collisions per unit transverse area, which is obtained from the
Glauber model calculation. As discussed above, the emerging hard partons from
the pQCD processes are then tracked through the plasma. It is assumed that no
interactions occur before the formation of the QGP at $\tau_i$, but thereafter
the local gluon density is updated in each time step taking the changes due to
expansion and the energy density profile into account. The procedure is stopped
when either the parton under consideration leaves the QGP or the local
temperature drops below $T_c$, the critical temperature of the phase
transition.

Hadronisation of partons emerging from the plasma presents new interesting
problems. The conventional models developed for $e^+e^-$ annihilation and
applied for $ep$ and $pp$, need not be applicable but there is little guidance
yet how the presence of a quark-gluon plasma affects the fragmentation of
energetic partons. It is also desirable to disentangle the effects of direct
energy loss and modified hadronisation. Therefore, in a first step, the
standard fragmentation procedure is used in this model. However, while the
standard Lund string fragmentation model \cite{Andersson:1983ia} is the best
option to use for the $pp$ reference, it is not easily applied to the case of
partons emerging from a plasma after several soft colour exchanges which have
changed the colour topology. It is quite unclear where the string from such a
parton should be connected and if the concept of the normal string applies at
all. The pragmatic way out that we have taken in this first study is to instead
apply independent hadronisation \cite{Field:1977fa,Sjostrand:1987xj}
for our simulations of AuAu collisions, but keeping the same basic 
fragmentation function $f(z)$ for the energy-momentum fraction $z$ given to 
the produced hadron in each iteration.

Finally, one also has to account for the Cronin effect, i.e.\ the $p_\perp$
broadening of the final state hadrons due to conventional initial state
scatterings. We have included this according to the model suggested in
\cite{Wang:1998ww,Zhang:2001ce}, 
i.e.\ the variance of the intrinsic $\kt$-distribution is increased by a
constant $\alpha$ for each scattering prior to the hard interaction,
i.e.\ $\sigma_{\kt}^2(x,y,b) = \sigma_{\kt,0}^2 + \alpha\cdot
(N_\mathrm{scat}(x,y,b)-1)$. In our model $\sigma_{\kt}$ does not depend on the hard
scattering momentum transfer scale $Q^2$ because parton showers are
treated explicitly. The parameter $\alpha$ was fixed independently with the
help of dAu data at the same energy.

\begin{table}[ht]
\caption{Parameters of the model for the quark-gluon plasma (using input from 
hydrodynamics and lattice calculations) and the SCI jet quenching mechanism.}
\label{tab_param}
\centering
\begin{tabular}{|l|l|l|l|}
\hline
\multicolumn{2}{|c|}{Parameter} & value & obtained from \\ \hline
QGP formation time & $\tau_i$ & \unit[0.2]{fm/c} & based on
saturation scale \cite{Eskola:1999fc}\\
energy density at $\tau_0 = \unit[1]{fm/c}$ & $\epsilon_0$ &
		$\unit[5.5]{GeV\,fm^{-3}}$ & fixed from hydro \cite{Huovinen:2002rn}\\
critical temperature & $T_c$ & \unit[0.175]{GeV} & fixed from lattice \cite{Karsch:2001vs}\\		
gluon mass & $m_g$ & \unit[0.2]{GeV} & chosen here\\
interaction probability & $P_\tn{int}$ & 0.5 & fixed from SCI \cite{Edin:1995gi,Enberg:2001vq}
\\
screening radius & $R_\tn{scr}$ & \unit[0.3]{fm} & cf.\ \cite{Karsch:2005ex}\\ 
width of Gaussian $t$ distr. & $\sigma_t$ & $\unit[0.5]{GeV^2}$ & chosen here \\
increase of $\sigma_{\kt}^2$ per scattering & $\alpha$ & $\unit[1]{GeV^2}$ & fitted from dAu data \\
\hline
\end{tabular}
\end{table}

The parameters of the model are listed in Table \ref{tab_param}. Not all of
them are free, the critical temperature $T_c$ is taken from lattice
calculations \cite{Karsch:2001vs}. The energy density has been determined with
hydrodynamic calculations \cite{Huovinen:2002rn} and can vary only in a small
range when considered at a fixed time $\tau_0=\unit[1]{fm/c}$. The formation
time is chosen in accordance with saturation scale considerations
\cite{Eskola:1999fc}. Using the known time evolution the energy density at any
time can be calculated, in particular at the time $\tau_i$ when the QGP is
formed. The soft interaction probability $P_\tn{int}$ is fixed from the SCI
model. The gluons in the plasma have an effective mass, which is a free
parameter. This mass has important dynamical consequences for the energy loss,
since a larger mass results in smaller energy loss. In fact, it was the
assumption of static scattering centres, i.e.\ infinitely massive plasma
partons, that gave vanishing energy loss through scattering so that only losses
through the medium-induced radiation was treated in \cite{Baier:1996kr}. The
width of the momentum transfer distribution $\sigma_t$ is free and may be
regarded as the most important parameter because it regulates how much energy
can be lost per collision.

The results are discussed in terms of the nuclear modification factor
and the two-particle azimuthal correlation. The nuclear modification
factor defined as 
\begin{equation}\label{eq:nuc-mod}
 R_\tn{AB}(\pt,\eta) = \left(\frac{1}{N_\tn{evt}} \frac{\d^2 N^\tn{AB}} 
   {\d\pt\d\eta}\right) \cdot \left( \frac{\langle N_\tn{coll} 
   \rangle}{\sigma_\tn{inel}^\tn{pp}} \frac{\d^2 \sigma^\tn{pp}} 
   {\d\pt\d\eta}\right)^{-1} 
\end{equation}
is a measure of deviations of the $\pt$-spectra obtained in
nucleus-nucleus collisions from the $pp$ result scaled with the average number
of binary collisions $\mean{N_\tn{coll}}$. The two-particle azimuthal
correlation 
\begin{equation}\label{eq:azimuth}
 D(\df) = \frac{1}{N_\tn{trig}}\frac{\d N}{\d (\df)} 
\end{equation}
is the azimuthal separation from a trigger particle, normalised
to the number of trigger particles in the data set. Here, both the trigger
particle and the other particles are required to be in certain ranges of $\pt$.

\begin{figure}[ht]
\centering
\input{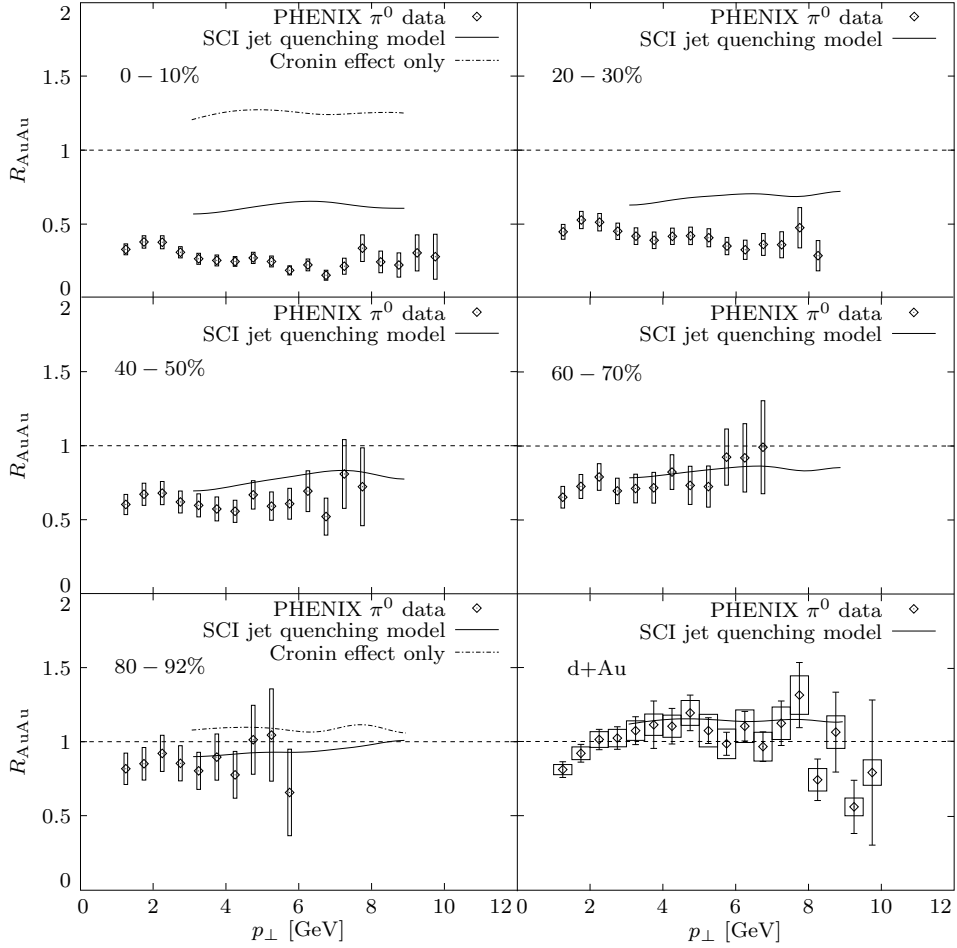}
 \caption{Nuclear modification factor, eq.~(\ref{eq:nuc-mod}), for different 
 centrality classes ($0-10$\% represents the most central and
 $80-92$\% the most peripheral collisions) of gold-gold (and minimum bias 
 deuterium-gold in lower right panel). PHENIX data \cite{Adler:2003qi} 
 compared to the SCI jet quenching model (starting at $\pt = \unit[3]{GeV}$ due
 to the cut-off used on \textsc{Pythia}'s hard scattering matrix elements). 
 The result from initial 
 state scattering, but no interactions with the plasma, is shown by the curves 
 'Cronin effect only'.}
 \label{fig_RAuAu}
\end{figure}

The model results for the nuclear modification factor (above \unit[3]{GeV}) are
shown in Fig.~\ref{fig_RAuAu} together with the PHENIX data \cite{Adler:2003qi}
for different centrality classes. Also shown are the results for d+Au
collisions that where used to determine the Cronin parameter $\alpha$. The
resulting Cronin effect of a broadened intrinsic $\kt$-distribution is shown
separately for the most central and the most peripheral AuAu collisions,
demonstrating that it is bigger in central events where the number of
collisions of a nucleon is larger for geometrical reasons.

\begin{figure}[ht]
 \centering
 \input{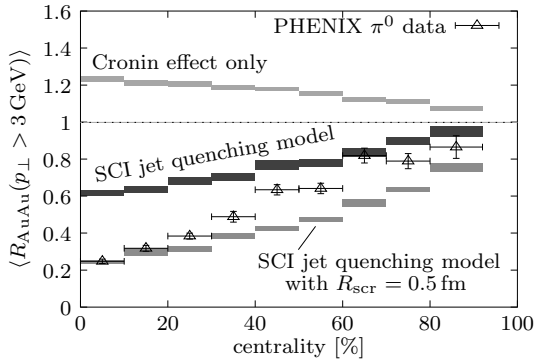}
 \caption{The nuclear modification factor $R_\tn{AuAu}$ (averaged 
 for $\pi^0$'s with $\pt > \unit[3]{GeV}$) versus centrality for AuAu data 
 \cite{Adler:2003qi} and for the SCI jet quenching model, with default
 parameters and with the screening radius $R_\tn{scr}$ increased from 0.3 to
 \unit[0.5]{fm}. The quenching effect should {\em not} be considered relative 
 to $R_\tn{AuAu}=1$ (dotted line) corresponding to no nuclear effects, but 
 relative to the upper curve including the Cronin effect.}
 \label{fig_cendep}
\end{figure}

The model reproduces the approximately flat shape of the $R_\tn{AuAu}$ data. 
Although the model is in agreement with the data in peripheral collisions, it
reaches only $\sim 50$\% of the suppression for the most central collisions.
The comparison of the SCI jet quenching model with the data in Figure
\ref{fig_cendep} illustrates again that the model is in agreement with the data
for peripheral and semiperipheral collisions but shows a weaker dependence on
centrality when going to more central collisions.  Nevertheless, even in the
most central collisions, the model shows a strong jet quenching effect which
should be taken relative to the increase in $R_\tn{AuAu}$ due to the Cronin
effect. It should be noted in this context that interactions with the
hadronic final state can also result in a significant suppression of high-$\pt$
hadrons \cite{Cassing:2003sb} and this effect would have to be added to any QGP
suppression.

In order to reproduce the data in central collisions with the SCI
jet quenching model alone one would have to roughly double the overall energy
loss. There are several ways how this can be achieved: One could for instance
increase the number of scatterings by increasing the screening radius from
\unit[0.3]{fm} to \unit[0.5]{fm}, or one could double the momentum transfer by
increasing $\sigma_t$ from $\unit[0.5]{GeV^2}$ to $\unit[2]{GeV^2}$. While with
the latter we think we are leaving the reasonable scale for soft interactions
one could consider the former, which effectively increases the  in-medium
parton-parton cross section from 3 to \unit[8]{mb}\footnote{In a recent study
the parton-parton cross section had to be increased to \unit[45]{mb} to
describe elliptic flow data in a parton-cascade \cite{Molnar:2001ux}}. It is
not sufficient to increase the number of scatterings by a factor 2 because the
energy loss in a single scattering decreases with decreasing parton energy so
that the parton loses less energy in the additional scatterings when
its energy is already reduced. With either of these two
parameter variations the SCI jet quenching model agrees with the data for the
most central collisions, but the centrality dependence is not linear so that
the calculation falls below the data in semi-central and, to a lesser degree,
peripheral events  (Fig.~\ref{fig_cendep}). 

The centrality dependence is sensitive to the energy and path length dependence
of the energy loss mechanisms and can be used to discriminate between different
types of model as investigated in \cite{toymodel}. In particular, models in
which the energy loss is based on coherent processes and grows with the path
length squared, seem to describe the linear centrality dependence better. 
However, also the energy dependence is relevant for the
dependence of $R_\tn{AuAu}$ on centrality. In the SCI jet quenching model,
using our default parameters, the energy loss in a single scattering is
negligible for small parton energies ($\lesssim \unit[1]{GeV}$ depending on the
plasma temperature), then it rises steeply and flattens at high energies
reaching roughly \unit[400]{MeV} at \unit[20]{GeV} in the case of light
quarks.

\begin{figure}[ht]
 \centering
 \input{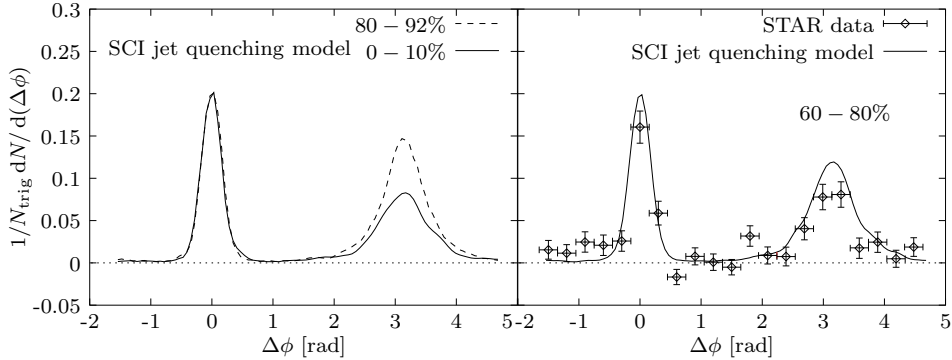}
 \caption{Model results for the 2-particle azimuthal correlation, 
 eq.~(\ref{eq:azimuth}), between a trigger particle (having $\unit[4]{GeV} < 
 \pt^\tn{trig} < \unit[6]{GeV}$ and defining zero azimuth, but not included in 
 the plot) and other particles (having $\pt > \unit[2]{GeV}$) 
 in peripheral ($80-92$\%) and central ($0-10$\%) collisions. The model
 is in reasonable agreement with the data for peripheral 
 collisions, but in central collisions the away-side peak disappears completely 
 in the data \cite{Adler:2002tq}.}
 \label{fig_d(deltaphi)}
\end{figure}

The resulting azimuthal correlation, Fig.~\ref{fig_d(deltaphi)}, show that in
peripheral collisions there is a clear jet-like peak at $\df = 0$ and a
somewhat lower and broader one at $\df = \pi$, much like in $pp$ interactions.
In central collisions, however, there is a clearly suppressed away side jet,
although the peak does not \emph{disappear} as
in the data \cite{Adler:2002tq,Adams:2003im}. To understand the quenching of
the away side jet is a general problem and we are not aware of any successful
model. Therefore, we discuss it here in more general terms. 

The amount of energy that a parton loses through the interactions with the QGP
is determined by the number of interactions and the energy loss in a single
scattering. The latter is dominated by the width of the $t$-distribution, but
depends also on the gluon mass and the temperature. A higher gluon mass or a
higher temperature lead to a smaller energy loss (the energy of the gluons in
the QGP is proportional to $T$ and, due to kinematics, the energy
loss is largest when the gluon has a small energy). While the QGP expands the
temperature drops according to $T \propto \tau^{-1/3}$ and the energy loss
mechanism becomes more efficient at later stages, but the density of gluons in
the plasma drops much faster, namely $n \propto \tau^{-1}$ so that the dilution
exceeds the effect of cooling. Furthermore the amount of energy lost by a
parton depends much stronger on the gluon density than on the temperature.
Consequently, the energy loss at early times dominates the total energy loss. 

\begin{figure}[ht]
 \centering
 \input{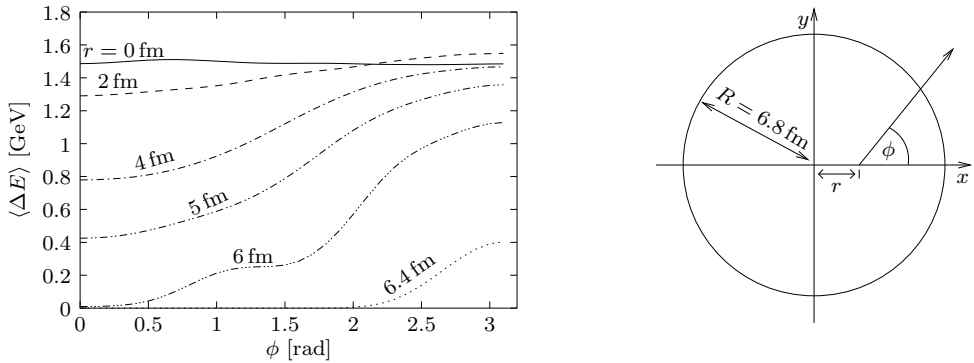}
 \caption{Energy loss $\Delta E$ of light quarks emerging from a hard scattering
  with \unit[5]{GeV} energy at different distances $r$ from the centre in central 
  AuAu collisions ($b=0$) with azimuthal angle $\phi$ (as illustrated to the 
  right). Quarks emitted at large $r$ with $\phi\to
  \pi$ have a long path through the collision region, but by the time they reach
  the centre (where the QGP is densest) the density has already dropped to very
  low values due to the expansion or the QGP has even already hadronised.}
 \label{fig_eloss} 
\end{figure}

Related to this is our finding that the largest energy loss is obtained when
the hard scattering takes place in the centre of the overlap region. This is
illustrated in Figure \ref{fig_eloss} which shows the energy loss of light
quarks in a central  collision for different emission points. When moving the
hard emission point from the centre towards the surface the energy loss gets
smaller because even those quarks that move towards the centre ($\phi=\pi$)
cannot have a longer {\em effective path length} (i.e.\ gluon
density integrated along the path) in the plasma than those
coming from the centre due to the expansion of the plasma. By the time a quark
coming from a distance ($r$) reaches the centre, the density has already
dropped and, moreover, for part of the quark's path-length there will even be
no plasma to interact with due to the plasma's limited life time of only
\unit[5.2]{fm/c} compared to the $\sim \unit[14]{fm}$ diameter of the gold
nucleus. 

Another point is that the locations of the hard interactions are, in fact, 
concentrated towards the centre of the QGP where the number of binary
nucleon-nucleon collisions per unit transverse area is highest. The path
lengths of the two hard-scattered partons are thus typically similar which
makes it unlikely that one of them loses much more energy than the other
(which is needed for the disappearance of one jet). Again, the limited plasma
life time (in relation to the gold nucleus size) prevents large asymmetries in
the path lengths through the plasma (the path length difference
cannot become larger than the plasma lifetime).

These three points prevent large asymmetries and may explain why it is so
difficult to get a substantial suppression (or disappearance) of the away-side
jet. On the other hand, the inhomogeneous energy density distribution can help
to amplify small asymmetries. Furthermore the hard interaction is at RHIC
energies dominated by $q+g \to q+g$ scatterings, which should also help because
the gluon interacts more strongly than the quark and should thus lose more
energy. These effects are included in our simulations, but are not strong
enough to create a large enough asymmetry with associated strong quenching
of the away side jet. 

It should be noted that this discussion is not only valid for this
particular model, but is of a more general nature since the
arguments are independent of the details of the interaction mechanism.
It should thus apply to all scenarios in which the energy loss depends
strongly on the density of scattering centres, which is also the case
for medium-induced gluon radiation.

In fact, it seems that the two-particle azimuthal correlation is sensitive to
details of the fragmentation procedure and is therefore afflicted with an
additional uncertainty. Already Lund string fragmentation and independent
fragmentation with the same fragmentation function lead to quite different
associated multiplicities. A softer fragmentation function produces more
hadrons but with lower momentum and since the parton $p_\perp$ spectrum is
steeply falling already small changes in the fragmentation function might cause
that the hadrons fall below the $\pt$ threshold of \unit[2]{GeV} used for the
azimuthal correlations.

In conclusion, energy loss due to soft scattering of energetic partons in the
QGP can contribute significantly to the jet quenching observed at RHIC. The SCI
jet quenching model gives a nuclear modification factor having the correct
$p_\perp$ dependence and a magnitude which can account for most of the effect
observed in peripheral collisions and about half the effect in central
collisions. There is also a suppression of the away-side jet, although not as
strong as in data. This depends on the distribution of hard
scattering events and initial energy density as well as the plasma evolution,
which are not specific to our particular model. This may give handles for
further investigations of the quark-gluon plasma. The centrality dependence of
the observed jet quenching may indicate the need for taking into account the
coherence between individual scatterings in the plasma. For an improved
understanding of the jet quenching phenomenon, one needs to take into account
both energy loss through medium-induced radiation and through scattering.

\end{document}